\documentclass{iau}
\usepackage{graphicx}

\title[O~{\small VI} from the SBs of the LMC]
{Observations of O~{\small VI} Absorption from the Superbubbles of the Large Magellanic Cloud}

\author[Pradhan et al.]
{Ananta C. Pradhan$^1$, Amit Pathak$^2$, Jayant Murthy$^3$
 \and D. K. Ojha$^1$}

\affiliation{$^1$Tata Institute of Fundamental Research, Homi Bhabha Road, Mumbai 400005, India, \\ email: {\tt acp@tifr.res.in} \\[\affilskip]
$^2$Department of Physics, Tezpur University, Tezpur 784028, India
 \\[\affilskip]
$^3$Indian Institute of Astrophysics, Koramangala II block, Bangalore 560034, India}

\pubyear{2013}
\volume{296}
\pagerange{xx--xx}
\jname{Supernova environmental impacts}
\editors{Alak Ray \& Dick McCray}
\begin{document}

\maketitle

\begin{abstract}
 We have presented the observations of O~{\small VI} absorption at 1032 \AA\ towards 22 sightlines in 10 superbubbles (SBs) of the Large Magellanic Cloud (LMC) using the data obtained from the {\em Far Ultraviolet Spectroscopic Explorer (FUSE)}. The estimated abundance of O~{\small VI} in the SBs varies from a minimum of (1.09 $\pm$0.22)$\times$10$^{14}$ atoms/cm$^{2}$ in SB N206 to a maximum of (3.71$\pm$0.23)$\times$10$^{14}$ atoms/cm$^{2}$ in SB N70. We find about a 46\% excess in the abundance of O~{\small VI} in the SBs compared to the non-SB lines of sight. Even inside a SB, O~{\small VI} column density (N(O~{\small VI})) varies by about a factor of 2 to 2.5. These data are useful in understanding the nature of the hot gas in SBs.

\end{abstract}

\firstsection 

\section{Introduction}
SBs are cavities across the interstellar medium (ISM) created by cataclysmic processes such as strong stellar winds and supernova explosions associated with massive stars in an OB association. The swept up material by stellar winds and supernova remnants produce an outer dense shell with an interior filled by low-density hot gas. The gas at the center of the SB is bright in X-rays, the outer shell formed by dust and cool gas is observed in the infrared while the interface between the cool shell and the hot interior is seen in the ultraviolet (UV). O~{\small VI} absorption lines at 1032 \AA\ and 1037.6 \AA\ are diagnostic of the energetic processes of the interface environments of the SBs which are produced by shock heating and are collisionally ionized. The LMC hosts more than 20 SBs which have been observed in X-rays and UV due to their proximity ($\sim$50 kpc). {\em FUSE} has observed O~{\small VI} doublet along many sightlines using early type stars as background objects. We have presented a study of O~{\small VI} absorption at 1032 \AA\ in the SBs of the LMC in order to assess the properties of intermediate environments in them.

\footnotesize{
\begin{table}
\caption{ Details of O~{\small VI} observations in the SBs of the LMC.}
\label{tab1}
\begin{tabular}{c c c c c c c c}
\hline
SBs      & Size of SBs              & {\em FUSE} Targets   & RA (J2000)   & Dec (J2000)    & FWHM       & Integration & N(O~{\small VI})/ \\
        &                         & 	     	   & hr min sec   & deg min sec    & (m\AA)     & limit(km/s) &  (10$^{14}$ atoms/cm$^{2}$) \\
\hline
N11     & \(10'\)$\times$\(7'\)    & PGMW-3070     & 04 56 43.25  &   -66 25 02.0  & 132$\pm$23 &  180, 345   & 1.27$^{+0.43}_{-0.43}$ \\
        &                          & LH103102      & 04 56 45.40  &   -66 24 45.9  & 132$\pm$23 &  180, 330   & 1.46$^{+0.10}_{-0.13}$ \\
        &                          & LH91486       & 04 56 55.58  &   -66 28 58.0  & 266$\pm$29 &  175, 385   & 2.96$^{+0.68}_{-0.68}$ \\
        &                          & PGMW-3223     & 04 57 00.80  &   -66 24 25.3  & 129$\pm$15 &  175, 315   & 1.27$^{+0.18}_{-0.18}$ \\
N51     & \(12'\)$\times$\(10'\)   & Sk-67D106     & 05 26 15.20  &   -67 29 58.3  & 180$\pm$7  &  175, 345   & 2.01$^{+0.50}_{-0.53}$ \\
        &                          & Sk-67D107     & 05 26 20.67  &   -67 29 55.4  & 254$\pm$10 &  160, 360   & 2.85$^{+0.27}_{-0.27}$\\
        &                          & Sk-67D111     & 05 26 47.95  &   -67 29 29.9  & 214$\pm$19 &  175, 365   & 2.20$^{+0.29}_{-0.24}$\\
N57     & \(12'\)$\times$\(7'\)    & Sk-67D166     & 05 31 44.31  &   -67 38 00.6  & 206$\pm$9  &  165, 390   & 2.09$^{+0.25}_{-0.20}$ \\
N70     & \(8'\)$\times$\(7'\)     & SK-67D250     & 05 43 15.48  &   -67 51 09.6  & 316$\pm$33 &  165, 375   & 3.71$^{+0.23}_{-0.23}$ \\
        &                          & D301-NW8      & 05 43 15.96  &   -67 49 51.0  & 228$\pm$30 &  175, 365   & 2.60$^{+0.09}_{-0.13}$ \\
        &                          & D301-1005     & 05 43 08.30  &   -67 50 52.4  & 284$\pm$57 &  165, 385   & 3.37$^{+0.17}_{-0.22}$ \\
N144    & \(13'\)$\times$\(12'\)   & HD36521       & 05 26 30.32  &   -68 50 25.4  & 126$\pm$9  &  175, 340   & 1.18$^{+0.28}_{-0.28}$ \\
        &                          & Sk-68D80      & 05 26 30.43  &   -68 50 26.6  & 303$\pm$16 &  145, 335   & 3.61$^{+0.30}_{-0.26}$\\
N204    & \(14'\)$\times$\(13'\)   & Sk-70D91      & 05 27 33.74  &   -70 36 48.3  & 256$\pm$7  &  160, 365   & 2.69$^{+0.24}_{-0.17}$ \\
N206    & \(9'\)$\times$\(15'\)    & BI184         & 05 30 30.60  &   -71 02 31.3  & 118$\pm$10 &  165, 330   & 1.09$^{+0.22}_{-0.22}$ \\
        &                          &  Sk-71D45     & 05 31 15.55  &   -71 04 08.9  & 194$\pm$9  &  160, 345   & 1.80$^{+0.24}_{-0.21}$ \\
N154    & \(12'\)$\times$\(8'\)    & SK-69D191     & 05 34 19.39  &   -69 45 10.0  & 185$\pm$25 &  165, 340   & 1.65$^{+0.26}_{-0.23}$\\
N158    & \(8'\)$\times$\(7'\)     & HDE269927     & 05 38 58.25  &   -69 29 19.1  & 245$\pm$8  &  160, 320   & 2.62$^{+0.16}_{-0.21}$ \\
30DOR C & \(7'\)$\times$\(6'\)     & MK42          & 05 38 42.10  &   -69 05 54.7  & 228$\pm$24 &  160, 330   & 2.60$^{+0.46}_{-0.45}$ \\
        &                          & SK-69D243     & 05 38 42.57  &   -69 06 03.2  & 307$\pm$15 &  150, 345   & 3.63$^{+0.40}_{-0.45}$ \\
        &                          & 30DOR-S-R136  & 05 38 51.70  &   -69 06 00.0  & 185$\pm$25 &  165, 320   & 1.77$^{+0.20}_{-0.24}$  \\
        &                          & SK-69D246     & 05 38 53.50  &   -69 02 00.7  & 211$\pm$7  &  155, 325   & 2.36$^{+0.18}_{-0.14}$ \\

\hline
\end{tabular}
\end{table}
}

\section{Measurement of O~{\small VI} Column Density}
We selected high resolution {\em FUSE} spectra from the archival data which were observed in the SB lines of sight in the LMC and had well pronounced O~{\small VI} profiles. We found 22 good O~{\small VI} observations covering 10 SBs in the LMC. The stellar continuum of O~{\small VI} profiles are of low order Legendre polynomials ($\le$ 5). The measurement of the optical depth of O~{\small VI} and the equivalent width for the LMC component are done by the apparent optical depth method following the procedures of \cite[Savage \& Sembach (1991)]{Savage91}, \cite[Sembach \& Savage (1992)]{Sembach92} and  \cite[Howk et al. (2002)]{Howk02}. The results are listed in Table \ref{tab1}. The 1$\sigma$ error in the measurements were estimated from the fitting procedures using the uncertainties in the {\em FUSE} data. The Milky Way ($v\lesssim$ 150 km/s) and LMC ($v\gtrsim$ 150 km/s) absorption components are well resolved by {\em FUSE} for most of the sightlines. The average value of the equivalent width of O~{\small VI} profiles of the SBs in the LMC comes out to be 214$\pm$19 m\AA.

\section{Results and Discussions}
N(O~{\small VI}) of SBs varies from (1.09$\pm$0.22)$\times$10$^{14}$ to (3.71$\pm$0.23)$\times$10$^{14}$ atoms/cm$^{2}$. The mean N(O~{\small VI}) for the SBs is found to be (3.07$\pm$0.62)$\times$10$^{14}$ atoms/cm$^{2}$ while the mean N(O~{\small VI}) for the non-SBs lines of sight is (2.10$\pm$0.50)$\times$10$^{14}$ atoms/cm$^{2}$ (\cite[Pathak et al. 2011]{Pathak11}). Thus, the SBs in the LMC show an excess O~{\small VI} abundance of about 46\% in comparison to non-SB regions. Studies for SB N70 (\cite[Danforth et al. 2006]{Danforth06}) found similar results with 60\% more O~{\small VI} than the non-SB targets. Considering N(O~{\small VI}) inside the individual SBs (e.g., N11 and 30 Dor C), a variation of about a factor of 2 to 2.5 in N(O~{\small VI}) was obtained. The thermal conduction between the interior hot, X-ray producing gas and the cool, photoionized shell of SBs is found to be the most favourable mechanism for the production of O~{\small VI} in the SBs (\cite[Danforth et al. 2006]{Danforth06}) as the thermal conduction models (\cite[Weaver et al. 1977]{Weaver77} and \cite[Borkowski et al. 1990]{Borkowski90}) have found N(O~{\small VI}) $\approx$ few $\times$ 10$^{13}$ atoms/cm$^{2}$, which is a close approximation of the observed N(O~{\small VI}). O~{\small VI} in the SBs does not show any correlation with the ISM morphologies such as H$\alpha$ and X-ray surface brightnesses. The temperature estimated from the line width of O~{\small VI} comes to be $\sim$10$^{6}$ K which represents slightly higher FWHM than expected for O~{\small VI} absorption.

\end{document}